\documentclass[runningheads, anonymous=true]{llncs}
\usepackage{graphicx}
\usepackage{xcolor}
\usepackage{amsmath}

\begin{document}
\title{Diagnosing Cardiac Abnormalities from 12-Lead Electrocardiograms Using Enhanced Deep Convolutional Neural Networks}
\titlerunning{Diagnosing Cardiac Abnormalities from 12-Lead ECGs}

\author{
	Binhang Yuan\inst{1}\and 
	Wenhui Xing\inst{2} \thanks{Corresponding author.}
}

\authorrunning{Yuan, B. and Xing, W. }

\institute{
Rice University, Houston, TX, USA\\
\email{by8@rice.edu} \and
Prudence Medical Technologies Ltd, Shanghai, China\\
\email{wenhui@prudencemed.com}
}
\maketitle              
\begin{abstract}
We train an enhanced deep convolutional neural network in order to identify eight cardiac abnormalities from the standard 12-lead electrocardiograms (ECGs) using the dataset of $14000$ ECGs. Instead of straightforwardly applying an end-to-end deep learning approach, we find that deep convolutional neural networks enhanced with sophisticated hand crafted features show advantages in reducing generalization errors. Additionally, data preprocessing and augmentation are essential since the distribution of eight cardiac abnormalities are highly biased in the given dataset. Our approach achieves promising generalization performance in the First China ECG Intelligent Competition; an empirical evaluation is also provided to validate the efficacy of our design on the competition ECG dataset.

\end{abstract}

\section{Introduction}

The electrocardiogram (ECG) is a diagnostic tool widely utilized for noninvasive diagnosis of various cardiovascular abnormalities in practice of clinical medicine worldwide. For example, there are approximately 250 million ECG recordings being processed by technicians for the diagnosis and treatment of patients with cardiovascular disease in China. The standard 12-lead electrocardiograms are the records of the heart's electrical activity collected from electrodes on arms/legs (known as limb leads) and torso (known as precordial leads). ECG interpretation plays an central role in the assessment of cardiovascular disease based on either a cardiologist's experience or computer-aided diagnosis systems. In practice, compute-aided interpretation has become increasingly important, since such technique improves the accuracy of diagnosis, facilitates health care decision making and reduces costs \cite{schlapfer2017computer}.

Traditionally, non-learning based approaches adopt wavelet, Fourier or other heuristic methods to classify specific abnormalities, but in practice these methods shows substantial rates of misdiagnosis \cite{shah2007errors}. Recently, deep neural networks (DNNs) \cite{goodfellow2016deep} have led to great success of machine learning to resolve diversified learning problems, for example, time series classification \cite{fawaz2019deep}. There has also been success in applying deep convolutional neural networks (CNN) to detect cardiovascular abnormalities from single-lead ECGs \cite{hannun2019cardiologist} or 12-lead ECGs \cite{hughes2018using}, where end-to-end CNN architectures are applied. On the other hand, we propose a novel deep neural network architecture enhanced by sophisticated hand crafted features, where such features are concatenated to the last fully connected layer in order to aid the activations from the original CNN to classify ECG signals. 

Another significant challenge is to train models on the highly heterogeneous ECG dataset. The distribution of cardiovascular abnormalities is highly biased in the dataset, due to the fact that some cardiovascular diseases appear at very low frequencies, while others are relatively common in population. Additionally, even within the ECG signals labeled as the rare abnormal class, the abnormal morphology only appears sporadically. The heterogeneity of the dataset makes sophisticated data preprocessing and augmentation essential to achieve good generalization performance. In attempt to address this issue, we adopt the follow tricks: i) the weights of different abnormal classes are adjusted according to their frequency; ii) the noisy ranges of ECGs are removed before applying the CNN model; iii) the signals are cropped according to a heuristic to augment the training dataset. 

The specific contributions of this paper are highlighted as follows: 
\vspace{-5pt}
\begin{itemize}
\item A deep CNN model augmented by domain specific features in order to both improve the ability of fitting the ECG training dataset, and reduce the generalization error on the test dataset;
\item A group of practical means of data preprocessing and augmentation in attempt to resolve the heterogeneity embedded in the ECG dataset;
\item An empirical evaluation to corroborate the effectiveness and efficacy of our architecture on the competition ECG dataset.
\end{itemize}
\vspace{-5pt}

In the end, we highlight that our approach achieves promising generalization performance in the First China ECG Intelligent Competition. 

\section{Dataset}

The dataset described in this paper is provided in the rematch stage of the First China ECG Intelligent Competition \cite{competition}. ECGs are collected with sample rate of 500 Hz and recorded in a standard 12-lead format including six limb leads (I, II, III, aVL, aVR and aVF) and six precordial leads (V1, V2, V3, V4, V5 and V6). The length of the ECGs varies from 4500 to 30000 representing the record in real time from $9.5$s to $60$s. There are $6500$ ECGs in the training dataset, $500$ ECGs in the validation dataset, and $7000$ ECGs in the test dataset. 

The goal of the learning application is to identify 8 cardiovascular abnormalities including atrial fibrillation (AF), first-degree atrioventricular heart block (FDAVB), complete right bundle branch block (CRBBB), left anterior fascicular block (LAFB), premature ventricular contractions (PVC), premature atrial contractions(PAC), early repolarization (ER),  T-wave changes (TWC) from the ECGs. Note that the labels of abnormalities are not mutually exclusive; in other word, a multi-label model is required to predict all possible diseases from an ECG. And if no abnormalities are detected, the model should indicate that this ECG is normal.

\section{The Architecture}

\begin{figure}
\centering
\includegraphics[width=0.45\textwidth]{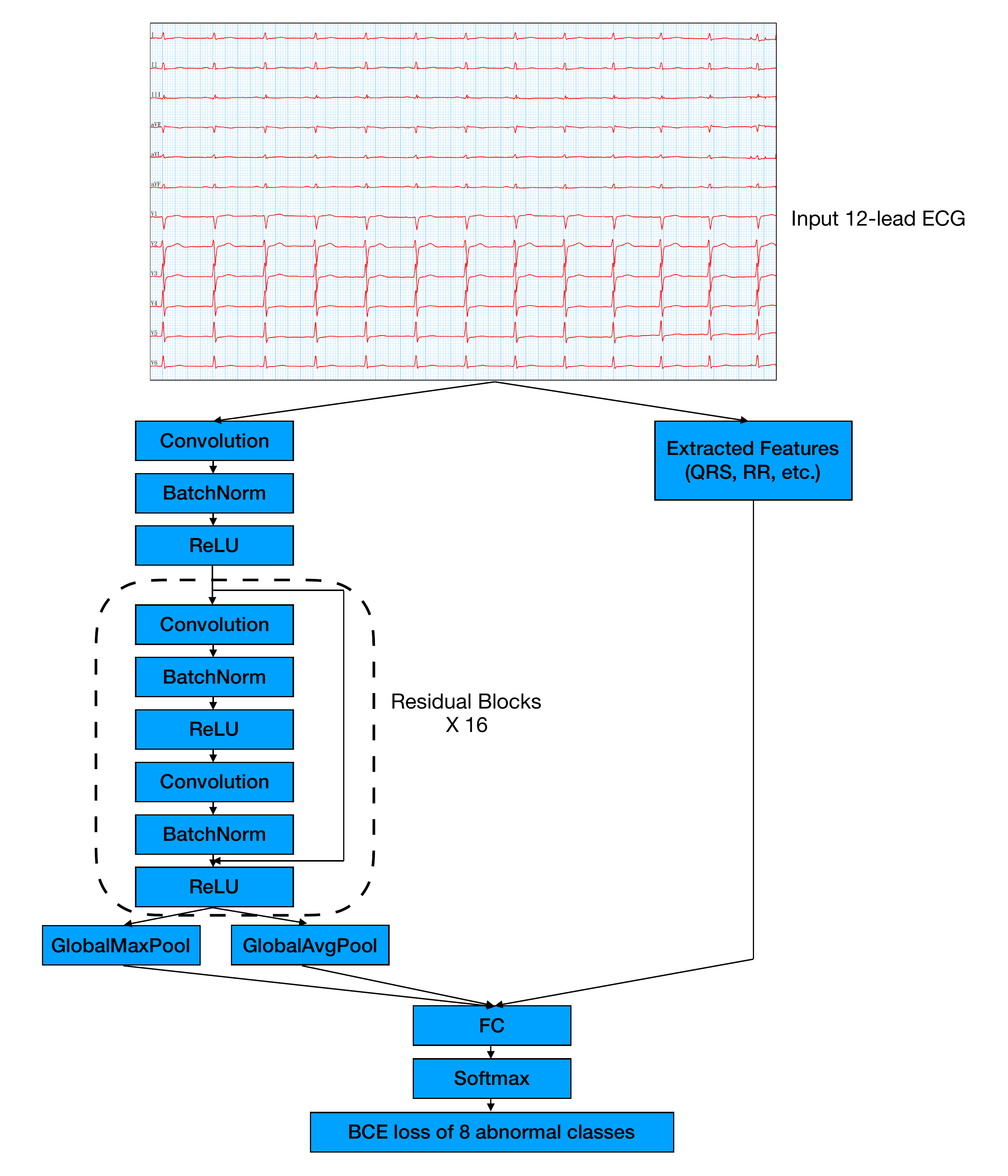}
\caption{\footnotesize The architecture of enhanced CNN. The CNN includes 16 residual basic blocks, where each residual blocks include 2 convolutional layers, 2 batch normalization layers \cite{ioffe2015batch}, and 2 ReLU activation layers \cite{nair2010rectified}. The output activations of the last residual block will be flattened by a global average pooling layer and a global max pooling layer. The concatenation of both global pooling layers and the abstracted features will be used by a fully connected layer. Lastly, the softmax function is applied before computing the binary cross entropy loss with the labels.} \label{fig1}
\end{figure}

The architecture of our network is summarized in Figure \ref{fig1}. The original 12-lead ECG signal is first denoised and then passed to the model. Heuristic features including QRS width, PR interval and standard deviation of the signal \cite{holm2010several} are abstracted from the ECGs on one side. On the other side, the ECG data batch goes through a 96-layer convolutional neural network with 16 residual basic blocks \cite{he2016deep}. Finally, the activations from the deep CNN will be flatten by global pooling layers, where the output will be concatenated with the heuristic features, and then will be utilized by a fully connected layer for the classifying task. Note that we encode a multi-label model where the parameters are shared for each abnormality instead of training 8 independent binary classifiers. We determine the layer number by choosing the shallowest network without compromising the generalization performance. Below we highlight a few important characters of our architecture. 

\subsection{Heuristic Features}

Traditionally, computer-aided interpretation systems substantially depend on the heuristic features, however, these straightforward approaches usually suffer from high misdiagnosis rates. On the other hand, these heuristic features materialize important domain-specific knowledge from the cardiologist's experience. For example, QRS width (in other term of QRS complex) represents the duration of three graphical deflections (known as Q wave, R wave, and S wave) on an ECG, corresponding to the depolarization of the right and left ventricles of the human heart and contraction of the large ventricular muscles; RR interval measures the time elapsed between two successive R waves of the QRS signal on the ECG, representing the intrinsic properties of the sinus node as well as autonomic influences. We find that in practice, end-to-end deep learning approaches can also benefit from such features. In fact, one can view the deep convolutional residual blocks as a magical black box to automatically extract features purely based on the input data and label. As a result, we combine such features with the heuristic features as the input for the last fully connected layer of the classifier in order to further reduce the generalization error.

\subsection{Global Pooling Layer}

Global pooling layers are widely used in CNN models, which flatten the activation tensors to two-dimensional before input the activation to afterwards fully connected layers. Different pooling layers tend to preserve different properties from the input activation \cite{boureau2010theoretical}. In general, max pooling takes the maximum activation in a block, which is good at retaining high frequency characteristics, while average pooling compresses the block by computing the mean of the block, which keeps the low frequency information. However, the detection of different abnormalities relies on different properties. For example, PAV and PVC only occurs in sporadic periods in the ECGs, which relies on infrequent properties, while other abnormalities (AF, FDAVB, CRBBB, LAFB, ER, and TWC) appears in every period in the ECGs. In order to get out of such dilemma, we compute both the global averaging pooling and global max pooling, and concatenate them for the later fully connected layer.

\section{Details of Learning}

In this section, we discuss the details of the learning procedure,  including data preprocessing and augmentation methods, and the optimization hyper parameters applied in the competition. 

\subsection{Data Preprocessing}

Since the abnormalities are substantially heterogenous from the competition dataset, we adjust the weights of each sample according to the frequency in the dataset so that the weighted loss function can emphasize the abnormal classes with less samples.
Additionally, the input ECGs are usually interfered by various noises e.g. power line interference, baseline drift, electrode contact noise \cite{joshi2013survey}, etc. We apply a wavelet based ECG denoising method \cite{khan2011wavelet} to remove the noises from the input ECGs efficiently. 

\subsection{Data Augmentation}
\begin{figure}
\centering
\includegraphics[width=0.8\textwidth]{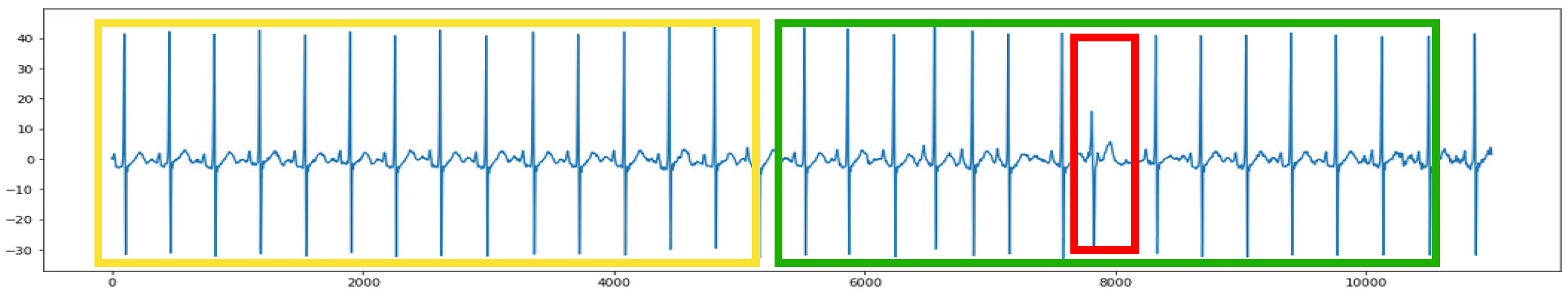}

\caption{\footnotesize An illustrative example of heuristic based cropping. The irregular range is marked by the red box, which potentially includes PAC or PVC. In the training phase, random cropping windows are generated, but only the window covers the marked region (eg., the green window above) are included in the training procedure, while the window excluding the marked range will be rejected (like the yellow box above).} 
\label{fig2}
\end{figure}

Importantly, the length of ECGs in the dataset varies in the dataset, naive approaches can be padding the short sequences with constant $0$, or stochastically cropping the signal to a fixed length. However, padding will include uninformative segments; while random cropping will take the risk of missing the informative parts in the original signal. For example, PAC and PVC only occurs in some periods in each ECG, if such region is not retained, the model can never learn the desired knowledge from the training data. 
In attempt to address this issue, we propose an innovative \textbf{heuristic based cropping} approach aimed at ECG dataset augmentation, where we first locate the QRS complex in the ECGs, according to the power spectrum and a group of bandpass filters \cite{kohler2002principles,pan1985real}; then the irregular QRS, T wave and P wave regions are marked as potentially problematic regions; finally only the cropping windows that include the potentially problematic regions for PAC and PVC are accepted, while other cropped ECGs will be rejected. We provide an illustrative example in Figure \ref{fig2}.

\subsection{Optimization}
The loss function we apply for optimization is the weighted binary cross entropy between the model's output and the label from the training set. We train our model for $70$ epochs using Adam stochastic gradient descent (SGD) optimizer \cite{kingma2014adam} with the batch size of $40$, the learning rate of $0.0001$ and the weight decay of $0.000001$. The learning rate is divided by $5$ when the validation error rate stops improving with the current learning rate. Note that we do not apply ensemble learning techniques to obtain the results reported in this paper, which in practice usually further improves the performance, while not preferred when proposing an innovative model.

\section{Results}
In this section, we will enumerate the detailed results.

\noindent
\textbf{Metric.} The evaluation metric in the competition is the average $F_1$ score for 9 labels (8 abnormalities and norm). To be more specific, suppose the true positive, false positive, true negative, and false negative counts for label $i$ are denoted as $TP^i$,  $FP^i$, $TN^i$, and $FP^i$, respectively, then the precision, recall and $F_1$ score for label $i$ are defined as:

\begingroup
\scriptsize
\begin{equation*}
{Precision}^i = \frac{{TP^i}}{{TP^i + FP^i}}
\end{equation*}

\begin{equation*}
{Recall}^i = \frac{{TP^i}}{{TP^i + FN^i}}
\end{equation*}

\begin{equation*}
F_{1}^{i} = \frac{2 {Precision}^i \dot {Recall}^i}{{Precision}^i + {Recall}^i}
\end{equation*}
\endgroup

And the evaluation metric is the average $F_1$ score:
\begingroup
\scriptsize
\begin{equation*}
{F_1} = \frac{1}{9}\sum\limits_{i = 0}^8 {F_1^i}
\end{equation*}
\endgroup

The final score our team archives during the rematch stage is \textbf{$0.879$} according to the above definition of the evaluation metric.

\noindent
\textbf{Detailed Experimental Comparison.} For the purpose of verifying our design, in Table \ref{table1}, we illustrate the incremental development of our approach with the techniques described in Section 3 and Section 4. We first try an end-to-end deep CNN approach (ResNet with the same residual blocks in Figure \ref{fig1}, but only including global average pooling before the final fc layer) where the $F_1$ score is $0.797$. Then we augment the CNN architecture with the ECG domain-specific features mentioned in Section 3.1 and increase the $F_1$ score to $0.832$. For above two benchmarks, random cropping is utilized to unify the signal length, and no data augmentation techniques are applied. Afterwards, we apply the heuristic data augmentation method introduced in Section 4.2 and improve the $F_1$ score further to $0.853$. Finally, with the same data augmentation technique, we switch the global average pooling layer before the last fully connected layer to the combination of global average pooling and global max pooling as we detailed in Section 3.2 and achieve the final $F_1$ score of $0.879$. 

\begin{table}[]
\centering
\begin{tabular}{|l|c|c|c|c|}
\hline
        & ResNet & Feature Enhanced & Data Augmented & Final Approach \\ \hline
Normal  & 0.835     & 0.873            & 0.900          & 0.914          \\
AF      & 0.902     & 0.950            & 0.951          & 0.962          \\
FDAWB   & 0.809     & 0.828            & 0.876          & 0.860          \\ 
CRBBB   & 0.992     & 1.000            & 0.992          & 1.000          \\ 
LAFB    & 0.842     & 0.812            & 0.889          & 0.944          \\ 
PVC     & 0.849     & 0.844            & 0.915          & 0.965          \\ 
PAC     & 0.625     & 0.776            & 0.860          & 0.874          \\ 
ER      & 0.480     & 0.522            & 0.412          & 0.500          \\ 
TWC     & 0.839     & 0.880            & 0.879          & 0.892          \\ \hline
\textbf{Average} & 0.797     & 0.832            & 0.853          & 0.879          \\ \hline
\end{tabular}
\caption{F1 score of incremental development in our approach.} \label{table1}
\vspace{-20pt}
\end{table}

\noindent
\textbf{Discussion.} There are a couple of interesting points we want to emphasize from the above experiments. Firstly, after adopting the ECG domain specific features, we find that the $F_1$ score for each abnormality increases in general comparing to the naive end-to-end CNN approach, which suggests that such domain specific knowledge does help to extract information difficult to learn by the deep residual blocks. In fact, recent research tends to include more traditional features from signal processing to improve the generalization performance of deep learning models on time series analysis. For example, \cite{zhang2017stock} includes frequency information by considering discrete Fourier transform to enhance LSTM \cite{hochreiter1997long}. 

Secondly, observe that there is a significant improvement of $F_1$ score for PVC and PAC after applying the heuristic based cropping technique for the training set augmentation, where PVC $F_1$ score increases from $0.844$ to $0.915$ and PAC $F_1$ score increases from $0.776$ to $0.860$. This confirms the speculation that simple preliminary detection of PVC and PAC abnormalities aids to generate training ECG samples with high quality, so that the learning process can be more effective. 

Lastly, the combination of global average pooling and global max pooling layers also introduces a general enhancement of the $F_1$ score for each abnormality. We ascribe this improvement to the fact that the combination of two global pooling layers preserves more information from the convolution channels. As one can imagine, global pooling dramatically compresses the activations from the preceding residual blocks, which unavoidably losses information. The combination of the two global pooling layers keeps both low frequency information from average pooling and high frequency characteristics from max pooling. As a result, the fully connected layer afterwards can utilize them to improve the prediction accuracy.

\section{Conclusion}

We recapitulate the main contributions of this paper, where a deep convolutional neural network enhanced by domain specific features is proposed to classify 8 cardiac abnormalities from 12-lead ECGs; practical data preprocessing tricks and a heuristic ECG data augmentation method are introduced to handle the noise and heterogeneousness in the ECG datasets; empirical evaluations corroborating the effectiveness and efficacy of our architecture are enumerated.

In the future, we plan to accumulate more ECG data to validate the robustness of our proposed model so that this technique can be deployed in real clinical settings.

\section*{Acknowledgement}Thanks to the committee for their great effort of organizing the First China ECG Intelligent Competition and the anonymous reviewers for their insightful feedback on earlier versions of this paper.

%
%

\bibliographystyle{splncs04}
\bibliography{main}

\end{document}